SCN1A/Na$_V$1.1 channelopathies: mechanisms in expression systems, animal models and human iPSCs models.

Massimo Mantegazza[1] and Vania Broccoli[2,3]

1 Université Côte d'Azur, CNRS UMR7275, Inserm, Institute of Molecular and Cellular Pharmacology (IPMC), 660 Route des Lucioles, 06560 Valbonne-Sophia Antipolis, France
Tel. +33 (0)493953425
Email: mantegazza@ipmc.cnrs.fr

2 San Raffaele Scientific Institute, Via Olgettina 58, 20132, Milan; Italy
3 National Research Council (CNR), Institute of Neuroscience, 20129, Milan, Italy
Email: broccoli.vania@hsr.it

Running title: Models of SCN1A/Na$_V$1.1 channelopathies.

Number of text pages 18
Number of words (summary) 184
Number of words (main text); 7620
Number of tables: 0
Number of figures: 1

- SCN1A/Na$_V$1.1 mutations are involved in different epilepsies and in familial hemiplegic migraine.
- We have reviewed pathological mechanisms identified with experimental models, highlighting advantages, limits and pitfalls of the models.
- Overall results point to Na$_V$1.1 loss-of-function and GABAergic neurons' hypoexcitability as the initial epileptogenic mechanism.
- Functional effects of migraine mutations are consistent with Na$_V$1.1 gain-of-function.
- Technical issues and pathophysiological remodeling generated more mechanistic complexity, which has still to be fully disentangled.

**Abstract**

Pathogenic *SCN1A*/Na$_V$1.1 mutations cause well defined epilepsies, including Genetic Epilepsy with Febrile Seizures Plus (GEFS+) and the severe epileptic encephalopathy Dravet syndrome. In addition, they cause a severe form of migraine with aura, Familial Hemiplegic Migraine. Moreover, *SCN1A*/Na$_V$1.1 variants have been inferred as risk factors in other types of epilepsy. We review here the advancements obtained studying pathological mechanisms of *SCN1A*/Na$_V$1.1 mutations with experimental systems. We present results gained with *in vitro* expression systems, gene targeted animal models and the iPSC technology, highlighting advantages, limits and pitfalls for each of these systems.

Overall, the results obtained in the last two decades confirm that the initial pathological mechanism of epileptogenic *SCN1A*/Na$_V$1.1 mutations is loss-of-function of Na$_V$1.1 leading to hypoexcitability of at least some types of GABAergic neurons (including cortical and hippocampal parvalbumin- and somatostatin-positive ones). Conversely, more limited results point to Na$_V$1.1 gain-of-function for FHM mutations. Behind these relatively simple pathological mechanisms, an unexpected complexity has been observed, in part generated by technical issues in experimental studies and in part related to intrinsically complex pathophysiological responses and remodeling, which yet remain to be fully disentangled.

Voltage-gated Na$^+$ (Na$_V$) channels are essential for neuronal excitability because they transiently increase the membrane conductance to Na$^+$ in response to depolarizations, initiating action potential generation[1]. In fact, upon membrane depolarization, they open (activate) within few hundred microseconds generating an inward, depolarizing Na$^+$ current. At depolarized potentials, they convert with kinetics of a few milliseconds to a non-conducting inactivated state, through a process termed fast inactivation. Often inactivation is incomplete, resulting in a small slowly-inactivating current with kinetics of tens of seconds, called "persistent".

Na$_V$ channels in the brain are formed by a principal α subunit (Na$_V$1.1 to Na$_V$1.9, coded by the genes *SCN1A* to *SCN11A*), constituted by four domains of six transmembrane segments each, associated with auxiliary β subunits (β1 to β4, coded by the genes *SCN1B* to *SCN4B*), which have a single transmembrane domain[2]. The α subunit forms the ion-conducting pore, is fully functional as Na$^+$ channel and is the target of antiepileptic drugs; β subunits modulate the functional properties and can be implicated in the subcellular targeting of α subunits[1; 3].

Na$_V$ channels have been implicated in numerous neurological diseases, and Na$_V$1.1/*SCN1A* is a major target of epileptogenic mutations[4].

**Pathogenic Na$_V$1.1/*SCN1A* mutations.**

The first Na$_V$1.1/*SCN1A* pathogenic mutations were identified by Escayg et al (2000): two missense mutations causing genetic epilepsy with febrile seizure plus (GEFS+)[5]. GEFS+ is characterized by febrile/ hyperthermic seizures, which often extend to adulthood, and large phenotypic variability extending to different types of epilepsy[6]. Dravet syndrome (DS) is a severe epileptic encephalopathy characterized by febrile/hyperthermic seizures at disease onset and later development of drug-resistant afebrile seizures, cognitive/behavioral deficits, ataxia and high mortality[7]. Because both DS and GEFS+ involve fever- and hyperthermia-induced seizures, the *SCN1A* gene was sequenced by Claes et al (2001) in DS patients, identifying 7 de-novo mutations[8]. Six of the mutations were predicted to produce a truncated nonfunctional Na$_V$1.1 protein, whereas the seventh was a missense mutation with unpredictable functional consequences, as for the GEFS+ mutations. Since these early studies, more than 1000 mutations have been identified in *SCN1A*. They cause more than 80% of DS cases, about 20% of GEFS+ cases and other rarer epileptic phenotypes; moreover, *SCN1A* genetic variants are risk factors for different types of common epilepsies[4]. A further well defined disease caused by missense *SCN1A* mutations is familial hemiplegic migraine type 3 (FHM3), a rare severe form of migraine with aura characterized by hemiplegia during the attacks[9]; in FHM3, seizures are very rare and not linked to migraine attacks[10].

Alike the first mutations described, Na$_V$1.1 mutations identified thereafter in DS are either missense (about one third), with functional effects that are difficult to predict, or truncations/deletions that are predicted to give rise to nonfunctional channels (about two-thirds). In contrast, GEFS+ and FHM *SCN1A* mutations are all missense. The identification of truncated, probably non-functional, epileptogenic Na$_V$ mutants was initially puzzling, because Na$_V$1.1 was thought to be expressed in excitatory neurons and its loss-of-function mutations were not consistent with neuronal hyperexcitability, which is thought to give rise to epileptic seizures. These puzzling findings highlighted the importance of functional studies for the identification of pathological mechanisms, which are still essential. In fact, although there have been progresses in the development of algorithms for *in silico* prediction of effects of mutations, in general based on conservation of protein's physico-chemical properties and of amino acid sequence within and across species [11], they cannot reliably disclose the detailed effect on protein's functions [12], and there are no tools for predicting overall effects on phenotypes.

**Functional effects on channel/neuronal excitability properties investigated in expression systems.**
Studies aimed at uncovering the functional effects of *SCN1A* mutations, in particular the missense ones, have been initially undertaken using heterologous expression in cultured cells that do not express the protein of interest, in which the cDNA (cloned DNA) of interest is inserted (transfected) [13]. They are in general human embryonic kidney (HEK) cells or the derived tsA-201 cell line, which allow relatively fast screens of mutants, but do not provide a neuronal cell background and do not form neuronal circuits for investigating effects on network activity. It is important to point out that the mutations should be introduced in the human cDNA of the isoform in which the mutation has been identified, because the same mutations may have different functional effects in paralog or ortholog sequences. However, the use of human Na$_V$1.1 cDNA has been technically challenging because of its instability leading to sequence rearrangements when handled with standard molecular biology techniques. In fact, although its sequence was identified in 1986[14], it was re-cloned and successfully expressed for the first time only sixteen years later, in a study aimed at identifying the functional effects of three GEFS+ mutations[15]. *SCN1A* undergoes alternative splicing (RNA processing of exons that generates proteins with different sequences from a single gene transcript) at exon 11, which produces variants with deletions of 33 base pairs (11 amino acids) and 84 base pairs (28 amino acids) within the intracellular loop between domain I and domain II; the -33-base pair shorter variant is the predominant splice variant expressed in brain [16]. Both full length and -33bp variants have been used for functional studies in expression systems.

**Epileptogenic mutations.** In the first functional study with human cDNA, Lossin and colleagues used the full length Na$_V$1.1 sequence and observed impaired inactivation resulting in increased persistent current for the GEFS+ mutants, a gain-of-function effect consistent with enhanced neuronal excitability, but opposite to the predicted effect of truncating DS mutations[15]. However, a follow-up paper reported loss-of-function for other GEFS+ mutations, in some cases complete loss-of-function, as for a DS missense mutation that was studied in parallel [17]. The functional investigations performed thereafter have confirmed that loss-of-function is the main effect of both DS and GEFS+ Na$_V$1.1 mutations, although for some mutations mixed loss and gain-of-function effects and for few mutations a net gain-of-function have been observed [18]. These incongruences can be generated by the experimental conditions: cellular background, cDNA splice variant, etc. For example, recordings from neocortical neurons dissociated from transgenic mice expressing the R1648H Na$_V$1.1 GEFS+ mutant, which is characterized by a gain-of-function in some heterologous expression systems[15], have shown that it induces instead loss-of-function in a neuronal cell background, with modifications of channel's properties that depend on the neuron subtype [19]. Another example is a study reporting that two mutants displaying gain-of-function at room temperature show instead loss-of-function at higher temperature [20].

The severity spectrum of Na$_V$1.1-related epilepsies could be a continuum and depend on the amount of loss-of-function of the mutant: a mild impairment of Na$_V$1.1 functions would cause mild phenotypes, whereas a more complete loss-of-functions would cause severe phenotypes[21]. Interestingly, some Na$_V$1.1 missense mutations cause loss-of-function because of folding/trafficking defects that lead to channel degradation in intracellular compartments [22]; these mutants can often be rescued by interacting proteins that probably stabilize the correct folding conformation [12; 23-25]. Similar interactions may rescue the mutants in vivo, regulate the amount of loss-of-function and modulate in this way the phenotype. Possibly, the complete loss-of-functions observed for some GEFS+ mutants may be caused by lack of rescue in the experimental conditions used and, conversely, lack of rescue *in vivo* could induce complete loss-of-function of DS mutants that show mild loss-of-function in some expression systems. Notably, folding defective Na$_V$1.1 missense mutants can also be partially rescued by interactions with small drugs or engineered peptides, which bind to them in the endoplasmic reticulum and probably stabilize the correct folding conformation, similarly to interacting proteins; this approach may be used for developing therapies [22].

**FHM3 mutations.** Functional studies of FHM3 mutations have initially generated similar incongruences. The first study was performed by engineering the identified mutation in the Na$_V$1.5 cDNA, the cardiac isoform, observing a mild gain-of-function effect [9]. The same mutation was then investigated with the long human Na$_V$1.1 splice variant in cell lines [26], observing mixed effects on gating properties that induced an overall loss-of-function, and, at the same time, in a study that used the shorter human Na$_V$1.1 splice variant (-11aa) expressed both in cell lines and in cultured neurons [27], observing numerous mixed effects that induced an overall large gain-of-function, leading to hyperexcitability of transfected neurons. Interestingly, another FHM mutation showed a nearly complete loss-of-function in one study [26], a puzzling effect that would be consistent with a severe epileptic phenotype, but a another study showed that the loss-of-function is caused by folding/trafficking defects and that the mutant can be rescued by interacting proteins or by expression in a neuronal cell background [28]. Strikingly, when partially rescued, this mutant showed a large gain-of-function because of the numerous modifications of gating properties, which induced hyperexcitability in transfected neurons. Notably, when this mutation was studied in the cardiac Na$_V$1.5 channel, it did not cause folding defects and induced mild modifications of functional properties that are consistent with moderate gain-of-function [29]. Other reports have now confirmed that FHM mutations cause gain-of-function of Na$_V$1.1 [30-32]. One of these reports identified another folding/trafficking defective FHM mutation for which the rescue induced by expression in neurons switched the functional effect from complete loss-of-function to large gain-of-function because of the modifications of gating properties[33]. Thus, this could be a recurrent mechanism in FHM3. Notably, a recurrent modification of gating properties induced by FHM3 mutations is the increase in persistent current. Na$_V$1.1 gain-of-function induced by FHM mutations may facilitate initiation and propagation of cortical spreading depression[27], which is a long lasting propagating depolarization of cortical circuits that is thought to be related to migrane aura, but this hypothesis has not been tested yet in animal models.

Overall, these results show that epileptogenic mutations cause a variable degree of loss-of-function whereas migraine mutations cause gain-of-function, which sometimes can appear as loss-of-function in functional studies because of rescuable folding/trafficking defects. They also highlight the importance of the cDNA variant and of the expression system in setting functional properties of Na$_V$1.1 mutants.

### Cellular/network mechanisms and phenotypes studied with animal models.

**Basic mechanisms in models of truncating mutations: hypoexcitability of GABAergic interneurons.** When the first *SCN1A* epileptogenic mutations were identified and functionally studied, available immunohistochemical data pointed to expression of Na$_V$1.1 mainly in somato-dendritic compartments of glutamatergic neurons[34]. Thus, it has initially seemed puzzling that mutations could lead to loss-of-function and reduced Na$^+$ current, consistent with reduced neuronal excitability, because epilepsy is a disorder characterized by brain hyperexcitability. This apparent incongruence was solved developing and studying gene targeted mice carrying *Scn1a* mutations. In the first study, Yu et al. (2006) reported the generation and investigation of a global knock-out (KO) mouse that modeled a truncating DS mutations[35]. Heterozygous knock-out (*Scn1a*$^{+/-}$ or Scn1a$^{tm1Wac}$) mice displayed seizures (including hyperthermia induced ones) and sporadic deaths beginning at postnatal day (P)21, with severity that was dependent on genetic background: very severe phenotype in the C57Bl/6 strain, mild phenotype in the 129 strain and intermediate phenotype in the mix C57Bl/6-129 F1. Importantly, it was observed that the Na$^+$ current density was reduced without modifications of gating properties in GABAergic interneurons, causing their hypoexcitability, but not in glutamatergic excitatory neurons. This suggested that the decreased excitability of GABAergic interneurons, induced by DS Na$_V$1.1 epileptogenic mutations, may cause reduction of GABAergic inhibition and network hyperexcitability: it was the first clear identification of the pathological mechanism of an epileptic encephalopathy. A further study with a knock-in model expressing a truncating non-sense DS mutation (*Scn1a*$^{R1407X/+}$ or Scn1a$^{tm1.1Kzy}$) reported similar phenotypic and cellular features, and showed that Na$_V$1.1 localizes to the axon initial segment of GABAergic interneurons, in particular parvalbumin (PV)-positive ones[36].

It has been hypothesized that DS Na$_V$1.1 truncating mutations cause haploinsufficiency: a 50% reduction of functional Na$_V$1.1 protein in heterozygotes, with complete loss-of-function and no effects on the wild-type protein. Recordings from dissociated neurons showed that half of the Na$^+$ current was lost in GABAergic neurons of *Scn1a*$^{+/-}$ mice, and a smaller additional decrease was observed in homozygous neurons[35; 37]. This nonlinear loss of Na$^+$ current has been hypothesized to depend on the compensatory upregulation of other Na$_V$ (e.g. Na$_V$1.3 upregulation has been observed in hippocampal GABAergic neurons of *Scn1a*$^{+/-}$ mice), which has been supposed to be larger in homozygous than in heterozygous neurons[35]. However, a dominant negative effect of truncated Na$_V$1.1 mutants could produce a similar nonlinear loss of current, because the Na$_V$1.1 current in heterozygotes would be reduced by co-expression of the mutant. Notably, experiments in expression systems have shown that truncating DS Na$_V$1.1 mutants do no induce negative dominance on wild type Na$_V$ expressed in brain, consistent with pure haploinsufficiency.

Several other subsequent studies, including those performed with conditional mouse models that allow the expression of mutations in specific neuronal subtypes (Scn1a$^{tm2.1Wac}$, Scn1a$^{tm2.1Kzy}$, Scn1a$^{Flox/+}$ [38]), have confirmed that hypoexcitability of GABAergic neurons is the initial pathological mechanism in DS models. In fact, it has been demonstrated that decreased excitability of GABAergic neurons actually leads to reduced GABAergic synaptic transmission[39], and that the specific deletion of Na$_V$1.1 in forebrain GABAergic interneurons is sufficient to induce a severe phenotype[40], similar to that of global DS mouse models, whereas the specific deletion in PV-positive ones induces a milder phenotype[38]. In fact, it has been shown that Na$_V$1.1 deletion impairs the excitability not only of PV-positive GABAergic neurons, but of at least also another GABAergic subpopulation: somatostatin-positive (SST+) neurons[41]. Notably, a study with conditional mouse models showed that Na$_V$1.1 loss-of-function in excitatory glutamatergic neurons has an ameliorating effect on the phenotype[42], consistent with the expression of Na$_V$1.1 in at least some subtypes of excitatory neurons, although reduced excitability of glutamatergic neurons has been reported only in one study and only at very high stimulation intensity[43].

Na$_V$1.1 begins to be expressed in mice at around P10 [36], but an overt epileptic phenotype, including hyperthermia-induced and spontaneous seizures, is observed after P20, identifying a pre-epileptic period in which the mutation is expressed but the modification induced in neuronal networks are not sufficient for generating seizures [44]. Notably, hippocampal hyperexcitability could be an important factor for seizure generation [44].

**Co-morbidities.** Gene targeted mice also replicate comorbidities identified in DS patients[7] (ataxia, cognitive behavioral deficits, sudden unexpected death in epilepsy-SUDEP- and sleep disturbances) and have been exploited for investigating their mechanisms.

Ataxia was the first co-morbidity investigated, reporting that global Scn1a$^{+/-}$ mice show motor deficits, including irregularity of stride length during locomotion, impaired motor reflexes in grasping, and mild tremor in limbs when immobile, consistent with cerebellar dysfunction; in fact, dissociated cerebellar Purkinje neurons (which are GABAergic) showed reduced Na$^+$ current and excitability[37].

Cognitive and behavioral deficits are main issues in the DS phenotype, often more disabling than seizures. Studies in different mouse lines, including conditional mice with selective deletion of Na$_V$1.1 channels in forebrain interneurons, reported hyperactivity, stereotyped behaviors, social interaction deficits, aversion to novel food odors and social odors, and impaired learning and memory[39; 45]. Some of these features have been interpreted as autistic-like behaviors, although clinical studies have reported that autistic traits are mild, if present at all, in DS patients, and social interaction deficits could be instead due to their motor and cognitive deficits[46; 47]. A follow up study used conditional Scn1a$^{+/-}$ mice to dissect the contribution of different subtypes of GABAergic neurons to cognitive and behavioral deficits[48], reporting that Na$_V$1.1 haploinsufficiency in PV+ interneurons causes deficits in social behaviors, but not hyperactivity, whereas haploinsufficiency in SST+ interneurons causes hyperactivity without deficits in social behaviors. Heterozygous Na$_V$1.1 deletion in both these interneuron types was required to impair long-term spatial memory in context-dependent fear conditioning, whereas short-term spatial learning or memory were not affected, consistent with the involvement of other types of GABAergic neurons. However, these results have been recently challenged by a study performed with a similar conditional model, reporting that $Na_V1.1$ haploinsufficiency in SST+ interneurons produces no noticeable behavioral anomalies, whereas haploinsufficiency in PV+ interneurons leads to hyperactivity, deficits in social behaviors, and cognitive decline; the authors hypothesized that this discrepancy might be linked to the different genetic background of the mice used in the two studies[49].

SUDEP occurs at higher rate in DS than in most other forms of epilepsy, and its mechanisms are not completely understood. Hypothetical mechanisms include cardiac dysfunctions, respiratory dysfunctions and cerebral shutdown during postictal depression of EEG activity (which is often observed after tonico-clonic seizures). Mortality rate is high also in DS mouse models and correlates with seizures severity. $Na_V1.1$ is expressed in the heart, in addition to the brain, and it has been reported that global $Scn1a^{R1407X/+}$ mice die after severe seizure-induced bradycardia and that their cardiomyocytes are hyperexcitable, which has been interpreted as indicating that intrinsic cardiac dysfunctions are the cause of the deaths[50]. However, selective knockout of $Scn1a$ only in forebrain interneurons in conditional $Scn1a^{+/-}$ mice results in both seizures and spontaneous death[40], whereas selective knockout in cardiomyocytes does not cause an overt phenotype[51]. Moreover, it was proposed that increased vagal tone induced by hypoexcitability of forebrain GABAergic neurons cause both post-ictal bradycardia and deaths, because they were reduced by selective blockade of peripheral muscarinic receptors. However, a more recent study challenged the vagal hyperactivity hypothesis, because it reported that DS patients show peri-ictal respiratory dysfunctions/prolonged postictal hypoventilation and that $Scn1a^{R1407X/+}$ mice die of central apnea followed by progressive bradycardia, which were not prevented by block of peripheral muscarinic receptors, whereas the block of central muscarinic receptors was effective[52]. The reason of the different effect of peripheral parasympathetic block in the two models is not clear yet. Another study, consistent with a central mechanism of cerebral shutdown, reported that in anesthetized $Scn1a^{R1407X/+}$ mice the epileptiform activity generated by topical application of the convulsant 4-aminopyridine to the cortex led to spreading depolarization in the dorsal medulla, a brainstem region that controls cardiorespiratory pacemaking, producing cardiorespiratory arrest[53]. Brainstem slices from $Scn1a^{R1407X/+}$ mice were more prone to the generation of spreading depolarization than controls, consistent with postictal brainstem spreading depolarization as a mechanism linking seizures and SUDEP in the presence of genetic mutations. However, it is not clear yet how cortical epileptic activity induces spreading depolarization in the brainstem and if this mechanism is at play also in awake animals.

Global *Scn1a*$^{+/-}$ mice show dysregulated circadian rhythms and sleep dysfunctions, whereas conditional *Scn1a*$^{+/-}$ mice with selective deletion of Na$_V$1.1 in forebrain GABAergic neurons show only sleep dysfunctions, consistent with involvement of different brain areas for the two defects[54]. Na$_V$1.1 is expressed in GABAergic neurons of the suprachiasmatic nucleus, the master circadian pacemaker that governs daily rhythms in mammals, and in global *Scn1a*$^{+/-}$ mice there is reduced activity of suprachiasmatic neurons and impaired ventro-dorsal communication within the nucleus, which may directly cause dysregulation of circadian rhythms[55]. Differently, it has been proposed that sleep dysfunctions are caused by defects of oscillatory activity in the thalamocortical network, because GABAergic neurons in the reticular nucleus of the thalamus show reduced Na$^+$ current and excitability (in particular reduced post-hyperpolarization rebound firing)[54]. Notably, this is a direct effect of the mutation and not a side effect of treatments, which is often hypothesized to be the cause of sleep dysfunctions in epileptic patients.

**Knock-in mouse model of the R1648H missense mutation.** In addition to models that carry truncating mutations, a knock-in mouse model of the missense mutation R1648H has been generated (*Scn1a*$^{R1648H/+}$ or Scn1a$^{tm1.1Aesc}$)[56], which is particularly interesting because R1648H shows large pleiotropy with phenotypes ranging from mild GEFS+ to DS [5; 57]. The mutation causes partial loss-of-function of Na$_V$1.1 because of gating modifications, as it was already reported with *in vitro* studies and transgenic mice (see above), and the phenotype of *Scn1a*$^{R1648H/+}$ mice is characterized by milder epileptic (hyperthermic and spontaneous seizures)[56] and sleep[58] phenotypes in comparison with mice carrying truncating mutations. Experiments in brain slices showed ubiquitous hypoexcitability of GABAergic interneurons in thalamus (in particular reduced post-hyperpolarization firing in the reticular nucleus, as later reported also in global *Scn1a*$^{+/-}$ mice, see above), cortex and hippocampus, because of deficit of action potential initiation at the axon initial segment, without detectable changes in excitatory neurons, leading to reduced action potential-driven GABAergic inhibition and increased abnormal spontaneous thalamo-cortical and hippocampal network activity[59]. Thus, mechanisms are similar to those observed in models of DS truncating mutations, and these mice show milder phenotype that recapitulate the mildest phenotypes in the GEFS+ spectrum.

**Modifiers and remodeling.** As already highlighted, the mouse genetic background has a strong effect on phenotype severity in *Scn1a* epileptic models. Studies have investigated the mechanism of genetic modifiers comparing different mouse strains. Using a further global knock-out (*Scn1a*$^{+/-TOT}$ or Scn1a$^{tm1Kea}$) model that completely eliminates Na$_V$1.1 expression removing the first *Scn1a* exon[60] (differently than the original *Scn1a*$^{+/-}$ model that expresses a truncated Na$_V$1.1 protein), it has been shown that *Scn1a*$^{+/-TOT}$ GABAergic neurons dissociated from 129 mice with mild phenotype have preserved Na+ current density, consistent with larger upregulation of $Na_V$ channels' expression in this strain, a compensatory mechanism probably linked to an intrinsic homeostatic response of GABAergic neurons. Interestingly, upregulation of $Na_V$1.3 in GABAergic neurons had been already observed in the first study of original global *Scn1a*[+/−] mice in the C57Bl/6 strain[35], but evidently in this strain the upregulation is less important. Another recent study performed with the global *Scn1a*[+/−] model showed that also excitability of GABAergic neurons recorded in brain slices is less impaired in 129 mice, probably because of homeostatic modifications of $Na_V$ channels' expression, but also because of different intrinsic properties of the neurons, disclosed comparing 129 and C57Bl/6 wild type mice[61]. More recently, further remodeling of gene expression has been reported comparing 129 and C57Bl/6 *Scn1a*[+/−−TOT] mice[62], including increased expression level of the GABA-A receptor α2 subunit, consistent in this case with a homeostatic response that is not intrinsic to GABAergic neurons. Notably, $Na_V$ genes were not identified as modifiers in this study.

A study of dissociated neurons from *Scn1a*[+/−−TOT] mice has reported hyperexcitability of pyramidal hippocampal neurons at the age in which animals show spontaneous seizures and mortality (>P20), which was not observed in the pre-epileptic period[60]. However, this result has not been confirmed in studies of brain slices. In fact, recordings in brain slices from *Scn1a*[+/−] mice, of the same age as the *Scn1a*[+/−−TOT] used in the previous study, did not disclose modifications in the firing properties of CA1 pyramidal neurons[61]. Moreover, a recent study[63] reported that the hypoexcitability of PV+ neurons in layer II/III of primary somatosensory cortex slices from *Scn1a*[+/−−TOT] mice is transient, because it was observed only before P35, but not after that age, probably because of remodeling of the axon initial segment; in this study, modifications of pyramidal neurons' excitability were not observed and properties of other interneurons or other brain areas were not investigated. Notably, the only investigation of the activity of GABAergic neurons in vivo was performed by juxtacellular recordings of cortical PV+ neurons from global *Scn1a*[+/−] mice in the pre-epileptic period, which did not show alterations, consistent with homeostatic responses[43]. However, discharge frequency of the recorded neurons was low, and alterations could be disclosed at higher firing frequency. A further study has reported that GABAergic neurons of the reticular nucleus of the thalamus in brain slices from *Scn1a*[R1407X/+] mice are instead hyperexcitable, showing increased post-hyperpolarization rebound firing because of remodeling leading to reduced expression of SK K+ channels, and that the pharmacological boost of the SK current reduces nonconvulsive seizures in these mice[64]. Although the authors of this study indicated that the observed dysfunctions of GABAergic neurons in the reticular nucleus of the thalamus are consistent with previous reports, other studies with *Scn1a*[R1648H/+ 59] and *Scn1a*[+/- 54] mice had instead previously reported the opposite effect: decreased post-hyperpolarization rebound firing, as outlined above. The reason for this discrepancy is not clear yet. A very recent study performed with *Scn1a*$^{R1648H/+}$ mice in a 50%-50% 129-C57Bl/6 F1 strain, which are asymptomatic and do not have spontaneous seizures, has shown that seizures induced in the period of seizure onset for *Scn1a* models can cause remodeling that leads to a severe DS-like phenotype [65]. The remodeling is dependent by the presence of the mutation (the same induced seizures have no effect in wild type littermates) and involve hyperexcitability of selective populations of excitatory neurons (e.g. hippocampal granular DG neurons become hyperexcitable, but CA1 pyramidal neurons do not). This result is not consistent with previous data obtained downregulating $Na_V1.1$ expression in the hippocampus of wild type mice by RNA interference that showed cognitive deficits without seizures[66]. However, RNA interference could induce a larger $Na_V1.1$ loss-of-function compared to gene targeted models that reproduce human mutations.

Overall, these results show that the initial pathological mechanism in mouse models of epileptogenic *SCN1A* mutations is $Na_V1.1$ loss-of-function and hypoexcitability of at least PV+ and SOM+ GABAergic neurons, observation that has been confirmed by numerous studies, those reviewed above as well as those performed with a further knock-out mouse model (Scn1a$^{tm1.1Swl}$)[67] and an ENU induced rat model[68]. This finding is also supported by an investigation performed in patients[69]. However, this initial dysfunction leads to and is accompanied by both homeostatic and pathologic remodeling, complex phenomena that depend on the type of cell, the age, the genetic background, the interaction between $Na_V1.1$ mutations and experienced seizures. We are beginning to shed light on this complex scenario, which could be similar for numerous diseases of the nervous system[70].

Our review is not focused on drug screens, but it is worth to mention that *Scn1a* mouse models, as well as simpler zebrafish models that allow larger screens of drugs, have also been used to test different therapeutic compounds and strategies, sometimes with significant amelioration of the phenotype[71].

## Modeling *SCN1A* epilepsies with iPSC technology

The generation and study of mouse models for *SCN1A* epileptogenic mutations has been crucial for understanding the neurobiological function of $Na_v1.1$ and the effect of its mutations, but it has also disclosed an unexpected complexity, as reviewed above. These results highlight the strong effects of different modifying factors on the dysfunctions of Na+ channels and on their effects on neuronal activity. Despite some important efforts, the exact molecular origin of these effects remains largely unknown, as pointed out above. Importantly, the inherent genetic complexity of this *SCN1A* disorders is evident in the human pathology as well, and unknown inherited factors can play significant effects in modulating phenotype-genotype correlations for a given *SCN1A* disease allele in humans.

To better elucidate these questions and generate human cell models, the development of the induced pluripotent stem cell (iPSC) technology has disclosed unprecedented opportunities. Until recently, experiments on human neurons were extremely limited by the accessibility to valid samples isolated from surgeries or postmortem tissue donations. Since the seminal discovery of iPSC reprogramming, adult human somatic cells can be reverted to a stem cell state and subsequently differentiated into neurons. In the last ten years, an increasing number of procedures to differentiate human iPSCs into different neuronal subtypes has been developed providing a flexible platform for an ease and rapid generation of human neurons with disease-causing mutations [72-74]. These methodological developments have provided the conditions to establish iPSCs from an increasing number of DS patients and generate human neuronal cultures to investigate the functional deficits and their underlying pathological mechanisms. Nonetheless, there are no studies of *SCN1A* FHM mutations performed with this technology.

The first results in modeling DS with iPSCs were reported by Liu and colleagues in 2013, through the generation and analysis of neurons from two patients with either heterozygous deletion or missense mutation in *SCN1A* and 3 undetermined control individuals by retroviral assisted integrating methodology[75]: This system relies on the random integration of the retroviruses expressing the reprogramming factors within the genome of the starter cells to be reprogrammed. Previous studies showed that multiple viral integrations are present in iPSCs obtained using this approach causing some risks of genotoxicity[76]. Patient and control iPSCs were then differentiated into forebrain neurons by embryoid-body generation, selection of neural progenitor cells (NPCs) and subsequent neuronal maturation up to 8 weeks. In these conditions, they described DS neurons increased repetitive firing compared to their counterparts derived from control neurons. They proposed that increased neuronal firing might lead to network hyperexcitability or increased synchronization sufficient to produce seizures and related neurological impairments. Thus, these effects would entail a cell-autonomous mechanism which compensate for the partial loss of $Na_V1.1$. Indeed, the authors reported that DS mutant neurons exhibited a significant increase in $Na^+$ current densities. These findings suggest the existence of a homeostatic mechanism to promote $Na^+$ current compensation, possibly by overexpressing other $Na_V$ isoforms. However, no changes in RNA expression levels were found in genes encoding for these channels, thus it was assumed that a post-transcriptional mechanism or the involvement of other ion channels might be responsible for this functional compensation[75]. At the same time, another study reported the generation of human DS neurons with iPSCs that were generated from patient's fibroblasts with integrating retroviral vectors expressing the four Yamanaka transcription factors[77]. iPSCs were then compared to control human embryonic stem cells (ESCs) and both cell types were differentiated in neuronal cultures highly enriched (~90%) in excitatory neurons for functionally studies. Interestingly, in line with the previous study, DS neurons developed sustained firing with larger and more action potentials than controls. In addition, voltage-clamp analysis revealed that Na$^+$ current inactivation was significantly slowed-down and incomplete, leading to large persistent current in DS neurons. These findings prompted the authors to suggest that the heterozygous *SCN1A* mutations analyzed in their study could act with a gain-of-function mechanism. Thus, both reports essentially converged in showing cell-autonomous hyperexcitability of DS neurons and, irrespective from the different differentiation procedures, the two studies collected comparable results in describing the abnormal activities of DS neurons. However, different pathological mechanisms underlying these defects were hypothesized by the two works: homeostatic over-compensation or mutant dominant effects. Nevertheless, these two mechanisms do not exclude each other or, alternatively, can be triggered by different mutations of Na$_v$1.1. In both studies, currents and membrane excitability were tested in iPSC-derived neurons differentiated for 4 to 8 weeks in culture. This is a rather short timing of differentiation for human iPSCs and might thereby model an early phase of neuronal maturation in vivo. In line with this argument, in both studies control iPSC-derived neurons exhibited short trains of action potentials confirming an inherent functional immaturity[75; 77].

**Promoting neuronal maturation and establishing molecular reporters for targeted electrophysiological recordings.**
Nearly at the same time, Higurashi et al. (2013) reported the generation of iPSCs from one healthy donor and one DS patient with very severe seizure clusters, profound cognitive disabilities and ataxia harboring a truncating mutation in SCN1A[78]. Patient and control iPSCs were then differentiated in neurogenic conditions stimulating the sonic-hedgehog pathway to induce an anterior ventralized neuronal phenotype corresponding to the GABAergic interneuron identity. Thus, the authors reported the generation of iPSC-derived neuronal cultures highly enriched in GABA+ neurons (~70%-90% depending by each cell line). For electrophysiological recordings the authors restricted their analysis to only iPSC-derived neurons with spontaneous resting membrane potential more negative than -40 mV, membrane capacitance between 30 pF and 70 pF and able to spike at least 10 or more action potentials. This *a priori* selection was a fundamental decision in order to select the most functional and mature neurons. In these conditions, current clamp analysis showed that patient-derived GABA+ neurons exhibited a reduced amplitude and failures in action potential generation particularly evident with high intensity stimulation. Altogether, these data led the authors to suggest that Na$_v$1.1 loss-of-function in patient-derived neurons caused their inability to sustain high frequency action potential discharges elicited by high injected current intensities. This phenotype is highly resembling the functional deficiency described earlier in PV+ and SOM+ GABAergic interneurons of *Scn1a* mouse mutants[35; 36], indicating a failure of the inhibitory neuronal activity as a primary cause of the disease. Although the authors could not score frank PV protein staining in their iPSC-derived neuronal cultures, they detected the presence of high levels of PV mRNA indicating the possible existence of PV+ neuronal precursors within the analyzed neuronal cultures. Indeed, PV protein levels strongly increase with postnatal development and might require cell intrinsic and non-intrinsic factors for their maturation. The results obtained by Higurashi and co-authors were then confirmed and further extended by more recent studies.

Sun et al. (2016) generated multiple iPSC lines from twins affected by DS and carrying a heterozygous missense SCN1A mutation (S1328P) that affects trafficking and causes loss-of-function in gating properties of Na$_v$1.1, as observed with *in vitro* studies in expression systems[79]. iPSC reprogramming was carried out by nucleofecting patient's fibroblasts with episomal plasmids expressing the reprogramming factors and, thereby, largely avoiding any genotoxic integration. The authors then defined the experimental procedures to promote the robust and reliable differentiation of iPSCs into telencephalic neurons with either an excitatory or inhibitory identity. To visualize either type of neurons and facilitate the selection of the cells to be recorded, they generated two lentiviruses with GFP fluorescent protein downstream to either the CaMKII or the Dlx1/2 promoter in order to activate the reporter in either excitatory or inhibitory neurons, respectively. Approximately 65% of the Dlx1/2-GFP labeled neurons were co-stained by anti-GABA antibody indicating that a large fraction of the fluorescent cells were inhibitory neurons. To confirm and further extend these results, cell soma of Dlx1/2-GFP positive neurons were isolated by suction through a glass pipette to detect expression of GABAergic markers by single-cell RT-PCRs. Interestingly, 70% of Dlx1/2-GFP positive neurons were expressing either GAD65 or GAD67 and only a small percentage expressed 5HT3aR, a serotonin receptor and specific marker of caudal ganglionic eminence (CGE)-derived inhibitory interneurons. In fact, more than 60% of the cells profiled expressed Calbindin, which is normally enriched in medial ganglionic eminence (MGE)-derived interneurons. This analysis enabled the authors to conclude that the iPSC-derived neuronal culture was enriched in MGE-derived interneurons, the class of neurons mostly affected by loss of Na$_v$1.1 in *Scn1a* mutant mice, including PV+ and SOM+ GABAergic interneurons (see above). Unambiguous identification of excitatory and inhibitory neurons is a powerful setting to perform selective electrophysiological recordings and distinguish the unique properties of either of the two neuronal populations. Using this approach, the authors could investigate the relative impact of the loss of Na$_v$1.1 in excitatory or inhibitory neurons derived from the same patient or control iPSC-derived neuronal cultures. For functional assessment, iPSC-derived neurons were recorded after 50 to 90 days of *in vitro* differentiation. This timing is comparable to the differentiation procedures used in the previous studies. However, in this last case immature human neurons were co-cultured with a rat cortical astrocyte monolayer, which are known to accelerate the maturation of human iPSC-derived neurons in culture and promote the acquisition of mature functional properties[80]. Thus, this is an advantageous condition to generate iPSC-derived neurons with a more advanced maturation stage in comparison with the previous works. Remarkably, voltage-clamp recordings showed that Na$^+$ currents were not different between control and patient CaMKII-GFP-labeled excitatory neurons, whereas a significant reduction in Na$^+$ current amplitude was observed comparing mutant and control Dlx1-2-GFP positive inhibitory neurons[79]. Consistent with these results, DS inhibitory neurons showed reduced firing with high current stimulation, and there was no statistical difference between control and DS excitatory neurons in the maximum firing frequency. These findings suggested that loss of the Na$_V$1.1 impaired the ability of only inhibitory neurons from DS to fire at high frequencies. Interestingly, the authors found that *SCN1A* mRNA levels were 1.6-5.6 higher in cultures of inhibitory neurons relative to excitatory neuronal cells derived from multiple control iPSC lines. Hence, these data indicate that Na$_v$1.1 expression is enriched in human inhibitory neurons and is detectable in iPSC-differentiated neuronal cultures, whose differentiation timing is much shorter compared to the natural time course of human brain development preceding disease onset. Altogether, these results indicated that Na$_V$1.1 is a major component of voltage-gated Na$^+$ currents in human inhibitory neurons and plays a pivotal role in controlling the excitability of these cells. However, Sun and colleagues analyzed DS neurons with a single mutation in SCN1A and therefore it remained unaddressed whether the impairment in iPSC-derived GABAergic neurons is differently modulated by distinctive Na$_v$1.1 mutations.

Kim et al. (2017) provided an initial response to this issue by generating iPSCs from two DS patients with distinct clinical manifestations and comparing their functional properties[81]. In fact, they selected the DS-1 patient that had significantly more severe symptoms respect to the DS-2 patient in terms of the frequency of seizures and the extent of intellectual disabilities. In line with previous results, iPSC-derived GABAergic neurons from both DS patients exhibited decreased Na$^+$ currents and action potential firing. In addition, numbers of APs elicited by current steps and the Na$^+$ current density of DS-1-derived GABAergic neurons were significantly more reduced than their counterparts derived from the DS-2 patient. Thus, the functional alterations observed in patient-derived neurons appeared to recapitulate the phenotype severity described in each donor. These results indicate that different *SCN1A* mutations produce a distinct impact on Na$^+$ current and excitability in iPSC-derived neuronal cultures. Hence, this work introduces the interesting possibility that patient-derived neurons may provide an *in vitro* system exploitable to predict phenotype severity of SCN1A-positive patients. However, additional studies modeling more Na$_V$1.1 mutations are necessary in order to strengthen this direct relationship.

**Dissecting Na$_V$1.1 function in iPSC-derived cardiomyocytes from Dravet patients.** iPSCs can be differentiated in principle in any somatic cell lineage to generate different cell types carrying the same gene mutation. Frasier and co-authors (2018) exploited the iPSC technology to generate patient and control cardiomyocytes (CMs) and evaluated their intrinsic properties, because *SCN1A* is expressed in both brain and heart and its mutations might affect as well cardiac excitability, possibly contributing to the increased risks of SUDEP in DS patients[82]. Interestingly, they found that patient-derived CMs showed heightened Na$^+$ currents and higher contraction rates compared to control CMs. Consistent with this observation, clinical monitoring of a DS patient, whose iPSC-derived CMs showed a significant increase in Na$^+$ currents, revealed abnormal T-wave inversions and lateral T-wave flattening in the electrocardiogram. Enhanced Na$^+$ currents might be explained by the consequent compensatory overexpression of other Na$_V$ in the cardiac tissue. Previous studies in DS mice showed different SUDEP mechanisms, with discrepancies that have not been completely clarified yet (see above). Results obtained in patient-derived CMs complement previous findings, suggesting that altered intrinsic cardiac excitability may be a features of DS patients and increase the risk of SUDEP. Importantly, this study corroborates that iPSC technology can produce a human cardiac cellular model for investigating SUDEP mechanisms caused by ion channel mutations.

**Limits and possible future developments of the technology**. In summary, the reported studies modeling DS with patient iPSCs were instrumental to describe functional deficits in mutant neurons that can be responsible to sustain epileptic hyperexcitability. However, different studies reported conflicting data, showing opposite changes in Na$^+$ currents and neuronal excitability in DS neurons derived from iPSCs. A first obvious possibility to explain this apparent controversy is that the different groups studied different sets of mutations which can differentially affect Na$_V$1.1 expression and function. In addition, some limitations and technical challenges intrinsic with the iPSC neuronal model might introduce important differences among the neuronal cultures analyzed in these studies. In fact, obtaining full maturation of differentiated iPSC-derived neurons remains an issue not easy to be obtained in vitro. This is particularly relevant for iPSC-derived inhibitory interneurons that experience important molecular and functional changes during postnatal brain development. It is likely that the diverging functional differences described in the two groups of studies were obtained using neurons at a different stage of maturation. On this view, the distinct functional changes described in these studies would reflect a different timing of neuronal maturation when DS and control human neurons were compared. Thus, earlier studies might have captured functional differences between DS and control neurons occurring only at a premature phase of human neuronal development. This particular functional state might be difficult to be recapitulated in animal models confirming the relevance of the iPSC-based system to uncover the complex developmental trajectory of human neuronal development.

In vitro neuronal maturation is a crucial aspect which should be taken with scrupulous attention when aiming at identifying subtle changes in neuronal functional properties. In fact, iPSC-derived neurons recapitulate human embryonic development, and their inherent immaturity poses an important hurdle to be seriously considered in functional studies. Neuronal differentiation by the dual SMAD inhibition protocol, co-culture with primary astrocytes and supplementation with neurothrophins together with a defined cocktail of small-molecules[83] are important requisites to accelerate neuronal maturation *in vitro* and obtain a more mature functional profile. Despite these important methodological improvements, the large functional heterogeneity between neurons in the same neuronal cultures will likely remain a relevant issue that should not be overlooked. This hurdle might be minimized using two approaches that are non-alternative to each other but rather complementary. First, functional readouts should be compared between neurons selected for equivalent intrinsic membrane properties in terms of resting potential and capacitance. Second, fluorescent reporters visualizing neurons with mature properties can be exploited for selecting cells for functional studies[84]. These advances represent crucial steps to select defined neuronal populations with comparable functional properties for unbiased disease mechanistic studies.

All the studies previously reported compared iPSCs derived from DS patients and healthy donors. However, individual differences that are independent by the disease but rather deriving from genetic background, age and environmental factors might act as strong contributors to cell variability and functional differences. Notably, it has been recently shown that remodeling in *Scn1a* mouse models can be induced by the interaction between the mutation and seizures[65]. Thus, hyperexcitability/epileptiform activity could modulate the properties of iPSC-derived neurons *in vitro* and should be monitored to compare cultures that experienced similar conditions. The use of multiple iPSC lines from different donors is a common way to minimize this genetic heterogeneity, although it cannot solve alone this issue. The generation of isogenic pairs of iPSC lines that differ only for the disease mutation represents an improved model that overcome confounding effects of individual variability. Isogenic iPSC pairs are commonly generated by employing site-specific nucleases for genome engineering aimed at introducing or correcting the disease mutation in an iPSC line of reference[85]. Alternatively, a study sought to generate isogenic lines exploiting the generation of iPSCs from peripheral blood cells of the asymptomatic individual carrying a truncating *SCN1A* mutation in mosaicism and mother of two brothers with DS [86]. Transgene-free reprogramming of peripheral blood lymphocytes generated 11 iPSC clones, two of which harbored the *SCN1A* heterozygous mutation causing DS in her children, whereas the remaining clones were wild-type. Isogenic iPSCs were then differentiated in neuronal cultures populated with different neuronal cell types. Interestingly, they found that mutant neurons exhibited increased mRNA and protein levels of tyrosine hydroxylase (TH) together with a heightened production of dopamine as quantified in the neuronal culture medium. These differences were not caused by an increased proportion of TH+ neurons in the mutant neuronal cultures. Thus, these results suggest the *SCN1A* mutation may be directly responsible for cell intrinsic changes in the dopamine biosynthetic pathway. The dopaminergic system is highly involved in the pathogenesis of autism-spectrum disorders, and these results raise the interesting hypothesis that altered dopamine signaling might contribute at least in part to the neurocognitive dysfunctions observed in DS patients. Functional studies of the dopamine system are warranted during the progression of the disease in DS patient to corroborate this hypothesis.

In summary, iPSC-based cellular systems provide an invaluable platform for modelling diseases caused by *SCN1A* mutations and defining their impact on human neuronal functions. The published studies are convincing examples that iPSCs can be differentiated in excitatory and inhibitory neurons and their activity can be assessed to define the downstream events triggered by *SCN1A* gene mutations. The new advances in culture conditions to coax iPSC differentiation into selective neuronal subtypes with enhanced neuronal functionality and connectivity will be extremely helpful to further validate the cellular consequences of $Na_V1.1$ loss. The current lack for an ease generation of different GABAergic populations with faithful properties (in particular of PV+ neurons with fast-spiking activity) remains a hurdle to properly model some aspects of *SCN1A* diseases. However, some important technical developments in this direction have been already reported[87-89], which might offer an improved system to re-evaluate some functional consequences of *SCN1A* mutations.

## Overall conclusions

Pathogenic *SCN1A*/$Na_V1.1$ mutations have been studied with several experimental models at different integration levels, obtaining important results about effects of mutations and pathological mechanisms at large (in particular the identification of loss-of-function of $Na_V1.1$ caused by epileptogenic mutations and leading to hypoexcitability of GABAergic neurons), as well as about the identification of possible therapeutic approaches. As implied by the definition of "model", there is no single model that can fully reproduce all the features of the original, in this case the human disease, and the choice of a model depends on the particular aspect under study, on the scale of the study (low- to high-throughput), as well as on the evaluation and minimization of technical pitfalls.

In vitro models allow higher throughput than in vivo ones and it would be important to use them for routine clinical diagnostic and stratification of patients. In particular, expression systems are relatively fast and cheap models, but we still need to identify and standardize correct experimental conditions; transfection of cultured neurons with the appropriate human clone may be a reliable model. The iPSC technology can potentially generate different human cells and allow important insights on patient-specific cellular dysfunctions, but it is still expensive and time consuming. Moreover, technical improvements are still needed to obtain the correct cell types with faithful mature intrinsic properties. In vivo systems, including rodents and zebrafish, can allow more integrated pathophysiological studies and drug screens evaluating different aspects of phenotypes. We have also begun to discover the complexity behind an initially relatively simple pathological mechanism, in which cell type-specific remodelling is implicated in different types of pathologic and homeostatic responses at different stages of the disease, which could be at play also in patients.

**Ethical publication statement**

We confirm that we have read the Journal's position on issues involved in ethical publication and affirm that this report is consistent with those guidelines.

**Disclosure**

None of the authors has any conflict of interest to disclose.

**Funding**

Work in Massimo Mantegazza's laboratory has been funded by the EC FP7 project 602531-DESIRE, the Investissements d'Avenir-Laboratory of Excellence "Ion Channel Science and Therapeutics" (LabEx ICST ANR-11-LABX-0015-01) and the ComputaBrain project (IDEX Jedi, Université Cote d'Azur); the laboratory is a member of the "Fédération Hospitalo-Universitaire" FHU-INOVPAIN.

Work in Vania Broccoli's laboratory has been funded by the EC FP7 project 602531-DESIRE and the Associazione Gruppo Famiglie Dravet, Italy.

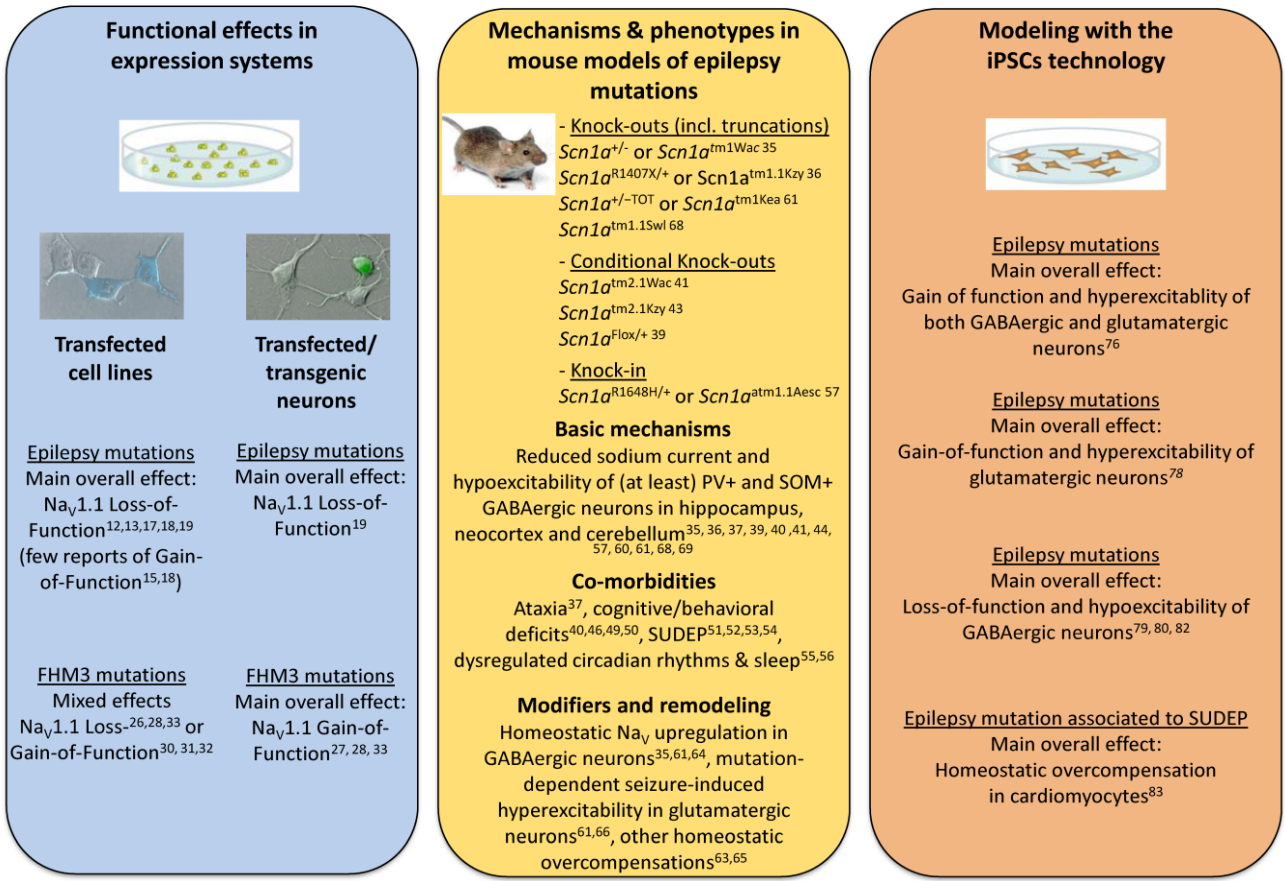

**Legend to figure**

**Schematic summary of the experimental studies of pathogenic Na$_V$1.1/SCN1A mutations**. For mouse models, both the names used in the review and the official names of the Mouse Genome Informatics database (http://www.informatics.jax.org/) are indicated. FHM3: Familial Hemiplegic Migraine type 3.